\documentclass[preprint,showpacs,prb,amsmath,amssymb]{revtex4}
\usepackage{graphicx}% Include figure files
\usepackage{dcolumn}% Align table columns on decimal point
\usepackage{bm}% bold math

\newcommand{\be}{\begin{equation}}
\newcommand{\ee}{\end{equation}}
\newcommand{\bea}{\begin{eqnarray}}
\newcommand{\eea}{\end{eqnarray}}
\newcommand{\bean}{\begin{eqnarray*}}
\newcommand{\eean}{\end{eqnarray*}}

\begin{document}
\title{Doped graphene as tunable electron-phonon coupling material} 

\author{Claudio Attaccalite$^{1,2}$, Ludger Wirtz$^3$, Michele Lazzeri$^4$, Francesco Mauri$^4$ and Angel Rubio$^{1,5}$ }
\affiliation{
$^1$ Nano-Bio Spectroscopy group and ETSF Scientific Development Centre, Dpto. Fisica de Materiales, Universidad del Pais Vasco, Centro de Fisica de Materiales  CSIC-UPV/EHU- MPC and DIP, E-20018 San Sebastian, Spain
\\
$^2$ Institut Neel, CNRS-UJF, Grenoble, France\\
$^3$ Institute for Electronics, Microelectronics, and Nanotechnology, CNRS-UMR 8520, Dept. ISEN, B.P. 60069, 59652 Villeneuve d'Ascq Cedex, France\\
$^4$ IMPMC, Universit\'es Paris 6 et 7, CNRS, IPGP, 140 rue de Lourmel, 75015 Paris, France\\
$^5$ Fritz-Haber-Institut der Max-Planck-Gesellschaft, Berlin, Germany}
\date{\today}                                         

\begin{abstract}
We present a new way to tune the electron-phonon coupling (EPC) in graphene
by changing the deformation potential with electron/hole doping. 
We show the EPC for highest optical branch at the high symmetry point K,
acquires a strong dependency on the doping level due to electron-electron correlation not accounted in mean-field approaches.
Such a dependency influences the dispersion (with respect to the laser energy) of the Raman D and 2D lines and 
the splitting of the 2D peak in multi-layer graphene. 
Finally this doping dependence opens the possibility to construct tunable electronic devices through the external control of the EPC.

%We present a new way to tune the electron-phonon coupling (EPC) in graphene 
%through the change of the deformation potential with electron/hole doping.  
%Calculations of the deformation potential on the level of density-functional
%theory with a (semi-)local exchange-correlation functional predict
%that the deformation potential is almost independent of the charge state.
%Taking electron-electron correlation into account, we show that for
%the highest optical branch at the high symmetry point K, the deformation
%potential is strongly dependent on the doping level. 
%This dependency influences the dispersion of the Raman D and 2D lines
%and in multi-layer graphene the splitting of the 2D peak.
%This make possible to measure the charge state of graphene via resonant Raman spectroscopy.  Finally this doping dependence opens the possibility to construct tunable electron devices through the external control of the electron-phonon coupling.  
\end{abstract}

%%      71.15.Mb   Density functional theory, local density approximation,
%%                 gradient and other corrections
%%      63.20.Kr   Phonon-electron and phonon-phonon interactions
%%      78.30.Na   Infrared and Raman spectra: Fullerenes and related materials
%%      81.05.Uw   Carbon, diamond, graphite

\maketitle

A large amount of work envisioning exciting new applications of 
graphene-based devices in nanoelectronics
has been published in the last years 
(see Ref.~\onlinecite{geim-rmp} and references therein). 
The performance of those electro-optical graphene-based-devices\cite{tsang}
is governed to a large extent by the electron-phonon coupling (EPC), or,
more precisely, by the deformation potential.
For example in high-current transport the scattering with phonons increases the differential resistance in carbon nanotubes and graphene \cite{zao,mauritube,highgraphene}.
It has been widely assumed that the deformation potential in graphene
is constant with respect to the electron/hole concentration.
For a proper description of device performances it is necessary
to control the validity of this approximation. Indeed, here we show
that the deformation potential displays a rather strong 
doping dependence which should be taken into account in the design
of new graphene devices.

The interaction between electrons and phonons in graphene and graphite has been 
studied with many experimental techniques ranging from Angle Resolved 
Photoemission Spectroscopy (ARPES)\cite{alexkc8}, inelastic x-ray 
scattering(IXS)\cite{alexnew},  Scanning Tunneling Spectroscopy (STS) 
\cite{andrei} to  Raman spectroscopy \cite{raman}. In particular, Raman 
spectroscopy
is commonly employed in graphene characterization because it is sensitive 
to the number of layers \cite{ferrari06,graf07}, the doping level 
\cite{pis07,efield, ferrari07b, stampfer07,baskonew,gpeakdoping} and the graphene 
edges \cite{baskoedge}. In order to interpret all above  mentioned 
experiments, a complete knowledge of the electronic structure, the phonon 
dispersion and the electron-phonon interaction is required.  
In graphene the  electron-phonon coupling(EPC) between the $\pi$ and 
$\pi^*$bands is responsible for the peculiar properties observed in the experiments \cite{calandra, drr, piscanec04}.

The dimensionless electron-phonon coupling for a mode $\nu$ at 
momentum {\bf q} is given by:
\bea
\lambda_{{\bf q}\nu} = \frac{2}{\hbar \omega_{{\bf q} \nu} N_\sigma (\epsilon_f) } \int_{BZ} \frac{d {\bf k} }{\Omega} \sum_{i,j} |g^\nu_{{\bf k} i,({\bf k+q})
j} |^2 \times \nonumber \\ 
\delta(\epsilon_k -\epsilon_f) \cdot \delta (\epsilon_{k+q} - \epsilon_f) 
\label{eq1}
\eea
where $\omega_{qv}$ is the phonon frequency, $N_\sigma (\epsilon_f)$ 
is the density of states per spin channel at the Fermi level, and  {\em i} and {\em j} are band indices.
The term
\begin{equation}
g^\nu_{{\bf k} i,({\bf k+q})j}  
= \langle {\bf k +q }, j | \Delta V_{{\bf q}\nu} | {\bf k}, i  
\rangle \sqrt { \hbar/(2 M \omega_q) }
\label{eq2}
\end{equation}
is the electron-phonon coupling matrix element that describes
the scattering of an electron from band {\em i} to band {\em j} due to
the phonon $\nu$ with wavevector {\bf q}.
The quantity $\lambda_{{\bf q}\nu}$ depends on the doping of the system
through the shift of the Fermi level and the subsequent change in
$N_\sigma (\epsilon_f)$. 
%Furthermore, it (weakly) depends on the doping through 
%the dependence of $g^\nu_{{\bf k} i,({\bf k+q}) j}$ on
%the phonon frequency which may change as a function of doping.
%For neutral graphene, $N_\sigma (\epsilon_f)=0$, and Eq.~\ref{eq1} is
%ill defined. 
Furthermore, it (weakly) depends on the doping through the variation
of $g^\nu_{{\bf k} i,({\bf k+q}) j}$. 
%Therefore, we discuss in the following the contributing
%matrix elements. Furthermore, in order to lift the dependence on the 
%phonon frequency, we calculate directly the matrix elements 
%$\langle {\bf k +q }, j | \Delta V_{{\bf q}\nu} | {\bf k}, i\rangle$.
In the following we are interested in the contribution coming from the matrix elements,
therefore, in order to lift the dependence on the phonon frequency, we calculate directly $\langle {\bf k +q }, j | \Delta V_{{\bf q}\nu} | {\bf k}, i\rangle$.
In particular, we will concentrate on the coupling of the $\pi$ and $\pi^*$
bands with the highest optical phonon branch (HOB) at ${\bf \Gamma}$ 
(E$_{2g}$ mode) and at {\bf K} (A$_1'$ mode).
We define 
\begin{equation}
\langle D^2_{\bf \Gamma}\rangle = \sum_{i,j}^{\pi,\pi^*}
\left| \langle {\bf K}, j | \Delta V_{{\bf \Gamma}E_{2g}} | {\bf K}, i \rangle
\right|^2/4
\end{equation}
and
\begin{equation}
\langle D^2_{\bf K}\rangle_{\bf k}^{\pi\pi^*} = \left|
\langle {\bf 2K + k }, \pi^* | \Delta V_{{\bf K}A_1'} | {\bf K+k}, \pi \rangle
\right|^2/2,
\end{equation}
where the sums are performed over the two times degenerate $\pi$ bands at {\bf K}.
In the limit of zero doping, $\langle D^2_{\bf \Gamma}\rangle$ is equal to $\langle D^2_{\bf \Gamma}\rangle_F$ 
as defined in Ref.~\onlinecite{lazzeri08}.
In the limit of zero doping and ${\bf k}\rightarrow {\bf 0}$,
$\langle D^2_{\bf K}\rangle_{\bf k}^{\pi\pi^*}$ is equal to $\langle D^2_{\bf K}\rangle_F$
as defined in Ref.~\onlinecite{lazzeri08}.
In fact, for small ${\bf k}$ the matrix elements between $\pi$ and $\pi$
(or between $\pi^*$ and $\pi^*$) are zero (see note [24] of Ref.~\onlinecite{piscanec04}).
In our earlier publications~\cite{lazzeri08} we called these quantities electron-phonon coupling.
In view of the above definitions (Eqs.~(\ref{eq1}) and (\ref{eq2}))
it is more precise to call $\langle D^2_{\bf \Gamma}\rangle$
and $\langle D^2_{\bf K}\rangle_{\bf k}^{\pi\pi^*}$ the ``square of the deformation potential''
(or, to be very precise, the ``average squared deformation potential of the 
$\pi$-bands'').
We focus our attention on $\langle D^2_{\bf \Gamma}\rangle$
and $\langle D^2_{\bf K}\rangle_{\bf k}^{\pi\pi^*}$, because these two quantities
are the ones responsible for the intensity and position 
of the peaks in Raman spectroscopy\cite{raman}, the kinks in ARPES and the 
phonon slope close to the Kohn-anomalies\cite{piscanec04}.

In graphene and carbon nanotubes, the deformation potential has
been usually obtained from tight-binding Hamiltonians computing the change 
in the nearest-neighbor hopping energy due to a lattice distortion 
\cite{ando}. In this approach the deformation potential, 
has been always considered a constant with respect to the electron or hole
density.  This approximation, although not justified microscopically, is widely used\cite{geim-rmp,pis07,efield,stampfer07,gpeakdoping,calandra,ando,dassarma}.
Moreover even {\it ab-initio} calculations using 
Density Functional Theory (DFT) in the local density
approximation (LDA) apparently confirmed that the deformation potential is weakly dependent on doping. 
However, including effects of electron-electron correlation, we will
show that 
for the {\bf K} (A$_1'$ mode) mode, this approximation breaks down
and  $\langle D^2_{\bf K}\rangle_{\bf k}^{\pi\pi^*}$ can change by more than $40\%$ just varying the electronic distribution (i.e. gated single and multi-layer graphene). 
This fact can be directly probed measuring the Raman D-peak dispersion 
of graphene versus doping as we will discuss in the following.

Recently it has been proven\cite{lazzeri08,basko} that in (neutral) graphene 
$\langle D^2_{\bf K}\rangle^{\pi\pi^*}_{\bf k=0}$ is strongly affected by electron 
correlation. In our previous work\cite{lazzeri08}, we have shown
that DFT-LDA or DFT-GGA underestimates $\langle D^2_{\bf K}\rangle^{\pi\pi^*}_{\bf k=0}$
by almost a factor of two. The electron-electron correlation  can be included on the
level of the GW-approximation obtaining a deformation potential which
reproduces the Raman D-line dispersion and the phonon slope around {\bf K} 
within few percent\cite{alexnew}. 
On the contrary, the deformation potential of the ${\bf \Gamma}$-E$_{2g}$ mode was shown to depend very little on electron-electron
correlation \cite{lazzeri08}. In this letter, we use the GW-approximation
to calculate the variation of the deformation potential with doping.

\begin{figure}
\centerline{\includegraphics[width=0.5\textwidth,angle=-90]{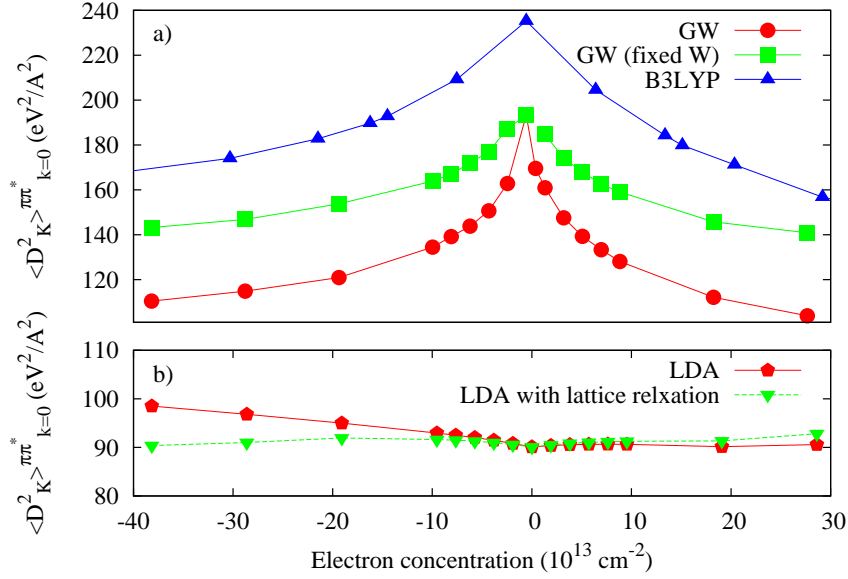}}
\caption{Panel $a$: squared deformation potential for the {\bf K}-A$_1'$  phonon between the $\pi$ bands 
$\langle D^2_{\bf K}\rangle^{\pi\pi^*}_{\bf k=0}$ in different approximations.
Circles are $G W$ results; squares are $G W(0)$ ones, keeping the screened potential $W(0)$ fixed to the undoped case;
triangles are B3LYP results. Panel $b$: variation of $\langle D^2_{\bf K}\rangle^{\pi\pi^*}_{\bf k=0}$
in LDA approximation, including or not lattice relaxation (triangles and pentagons, respectively).
\label{lda}}
\end{figure}

Doped graphene can be created in single layer field-effect transistor (FET) 
based experiments %\cite{fet1,fet2},
where an electron concentration up to $3\cdot10^{13}$ cm$^{-2}$ electron 
can be realized, while higher dopings are obtained using intercaled 
graphite \cite{kc8}.
In order to simulate doped graphene we employed a slab-geometry, 
i.e., bulk geometry with large distance
between the layers, changing the number of electrons in the unit cell 
and then compensating the negative/positive charge
with a uniform positive/negative background, see Ref.~\onlinecite{calculation} 
for details. 
The electronic and phonon structure of graphene at different doping levels 
were computed using DFT-LDA. The deformation potential is obtained using 
the scheme proposed in Ref.~\onlinecite{lazzeri08} based on a 
{\it frozen-phonon} approach by looking at the modification of the electronic 
structure upon displacement of the atoms following a given normal mode. 
The major advantage of this approach is that it can be used  with electronic
structure methods other than DFT. When DFT is used, this approach
gives the same result of density functional perturbation theory(DFPT).
First of all, we investigated the effect of the change in the
lattice constant  {\emph a}, induced by the doping, on the deformation 
potential. Using the functional dependence of {\emph a} versus the electron 
concentration from Eq.~(2) of Ref.~\onlinecite{laz06}, we calculated the 
deformation potential for different doping level with the corresponding 
lattice parameters.  In panel b of Fig.~\ref{lda} we compare the deformation 
potential calculated with and without lattice relaxation for different 
electron/hole doping. The difference between the two results is small 
when compared with renormalization effects that we are going to describe
below. Therefore, in order to make the analysis simpler, we performed all 
the calculations with the graphene experimental lattice constant.

On the level of the LDA, the variation of $\langle D^2_{\bf K}\rangle_{\bf k}^{\pi\pi^*}$ 
and $\langle D^2_{\bf \Gamma}\rangle$ with doping is very small.
In the following, we will introduce correlation effects beyond DFT-LDA on the deformation potential. 
We included these effects in $GW$ approximation \cite{GW} that has been successfully 
applied in the study of graphene and graphite\cite{louie,tbarpes} and its compounds\cite{zhou,graphite}. 
First of all we studied  the quasiparticle band-structure versus doping (see also Refs.~\onlinecite{polini} and \onlinecite{attaccalite}), that will be subsequently used to calculate the D-peak dispersion.  In table \ref{table1}, we report the change in the gap between the $\pi$ and $\pi^*$ at the high symmetry point {\bf M}, $\Delta \epsilon_M$, as function of doping. This quantity is directly related to the optical properties 
of graphite and graphene-based materials, and provides an alternative way to measure quasi-particle renormalization effects.  The strongest renormalization effect (compared to the LDA-gap) is present for zero doping. Electron/hole doping rapidly decreases the GW renormalization of the quasi-particle band structure. The same is true for the Fermi velocity $v_F$. Then from the change in the quasi-particle band structure upon atomic displacement, we calculated 
$\langle D^2_{\bf K}\rangle_{\bf k}^{\pi\pi^*}$ in the same way as was done for LDA (see ref. \onlinecite{lazzeri08}).
%Now, we discuss the effect of doping on the deformation potential. 
We found that for $\langle D^2_{\bf \Gamma}\rangle$ the GW result of 
ref.\onlinecite{lazzeri08}  is mainly unaffected by the doping 
level\cite{nonadiabatic}. The situation is completely different for 
$\langle D^2_{\bf K}\rangle^{\pi\pi^*}_{\bf k}$. 
In Fig.~\ref{lda}, and table \ref{table1}, we report the value of 
$\langle D^2_{\bf K}\rangle^{\pi\pi^*}_{\bf k=0}$ as a function of the doping in 
different approximations. In LDA (panel b) it is almost a constant, 
but the situation is completely different at the GW level 
(panel a):
$\langle D^2_{\bf K}\rangle^{\pi\pi^*}_{\bf k=0}$ is increased more than $80\%$ with
respect to the LDA result (at zero doping) and it acquires a strong doping 
dependence. The renormalized deformation potential rapidly decreases with 
electron or hole doping and gets close to the LDA value at large doping.  
%The result is symmetric for small electron/hole doping due to the graphene 
%band structure, but this symmetry is lost for larger doping.

\begin{table}
\centering
\caption{Squared deformation potential for the {\bf K}-A$'_1$ phonon versus doping, in $G W$ approximation.
$\langle D^2_{\bf K}\rangle^{\pi\pi^*}_{\bf k=0}$ and $\Delta\epsilon_{\bf M}$ are defined in the text, the LDA value of $\Delta\epsilon_{\bf M}$ at zero doping is $4.01115eV$. 
D-slope is the slope of the Raman D-peak dispersion (peak energy versus laser frequency), see the text.
\label{table1} } 
\label{tab1}
\begin{tabular}{c|cccc}
      &\multicolumn{4}{c}{Graphene:}\\
$\Delta n$ & $\langle D^2_{\bf K}\rangle^{\pi\pi^*}_{\bf k=0}$ & $\Delta\epsilon_{\bf M}$  & D-slope & \\
$10^{13} cm^{-2}$ & (eV$^2$/\AA$^2$)                     & eV                      & $cm^{-1}/eV$ \\
\hline
-38.16         & 110 & 4.550 & 27.58 \\
-28.62         & 115 & 4.585 & 29.83 \\
-19.08         & 121 & 4.632 & 32.41 \\
-9.54          & 135 & 4.711 & 37.84 \\
-7.63          & 139 & 4.723 & 39.49 \\
-5.72          & 144 & 4.761 & 41.20 \\
-3.81          & 151 & 4.795 & 43.67 \\
-1.90          & 163 & 4.820 & 48.20 \\
 0.00          & 193 & 4.867 & 58.40 \\
 1.90          & 161 & 4.803 & 47.82 \\
 3.81          & 148 & 4.765 & 43.26 \\
 5.72          & 139 & 4.724 & 40.24 \\
 7.63          & 133 & 4.667 & 38.44 \\
 9.54          & 128 & 4.642 & 36.75 \\
 19.08         & 112 & 4.532 & 31.53 \\
 28.62         & 104 & 4.485 & 29.11 \\
\end{tabular}
\end{table}   

In the following, we discuss the origin of the strong doping
dependence of $\langle D^2_{\bf K}\rangle^{\pi\pi^*}_{\bf k=0}$ within
the GW-approximation. Both the Green's function, $G$,
and the screened Coulomb potential, $W$, are doping dependent.
In order to disentangle the two effects, 
we performed test calculations within the GW approximation
keeping the screened Coulomb interaction fixed to its value at zero doping. 
The result is shown in Fig.~\ref{lda}. 
The doping dependence of the deformation potential is reduced by
about a factor of two. Therefore, we conclude that
the role of the screening is comparable to the correction coming from 
the Green's function variation. In fact the shift of the Fermi level due
to doping, leads to a suppression of transitions 
from $\pi$ to $\pi^*$ states. This affects the deformation potential 
both through a change of the screening and of the Green's function.

\begin{figure}[h]
\begin{center}
\includegraphics[width=0.5\textwidth,angle=-90]{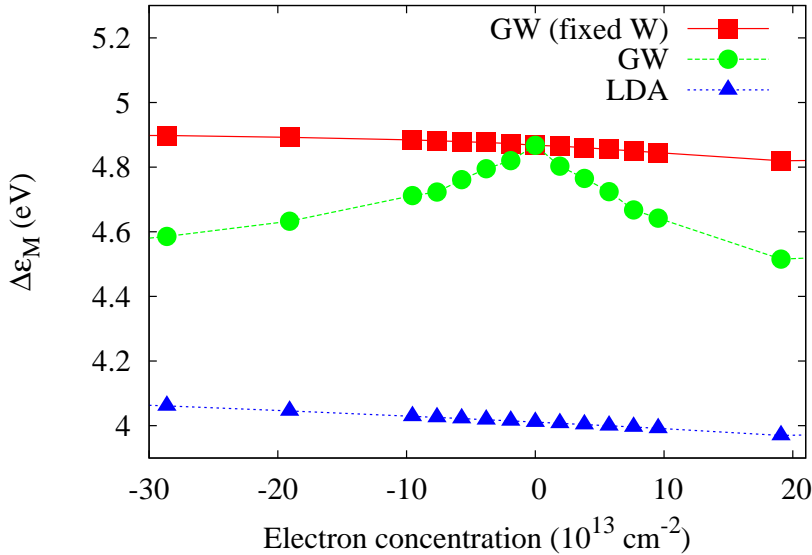}
\caption{Electronic gap at $M$ point ($\Delta\epsilon_{\bf M}$) as function of doping in different approximations: GW with fixed
screened Coulomb interaction, GW, and LDA \label{m_gap}}
\end{center}
\end{figure}

We compare the result for deformation potential with the one of the $\Delta \epsilon_M$ gap. For the latter we found that all the doping dependence is due to the screened Coulomb potential and not due to the change in the Green's function, see fig. \ref{m_gap}, while in the first both effects contribute to the its behavior with doping. This is due to the fact that, for the {\bf K}-A$'_1$ phonon, GW not only renormalizes the bare Green function lines but also introduces vertex corrections that behave different with the doping. However in general this is not true for other phonon modes, for instance for the ${\bf \Gamma}$-E$_{2g}$ mode vertex corrections have been proven to be negligible\cite{basko}.

\begin{figure}[h]
\centerline{\includegraphics[width=0.5\textwidth,angle=-90]{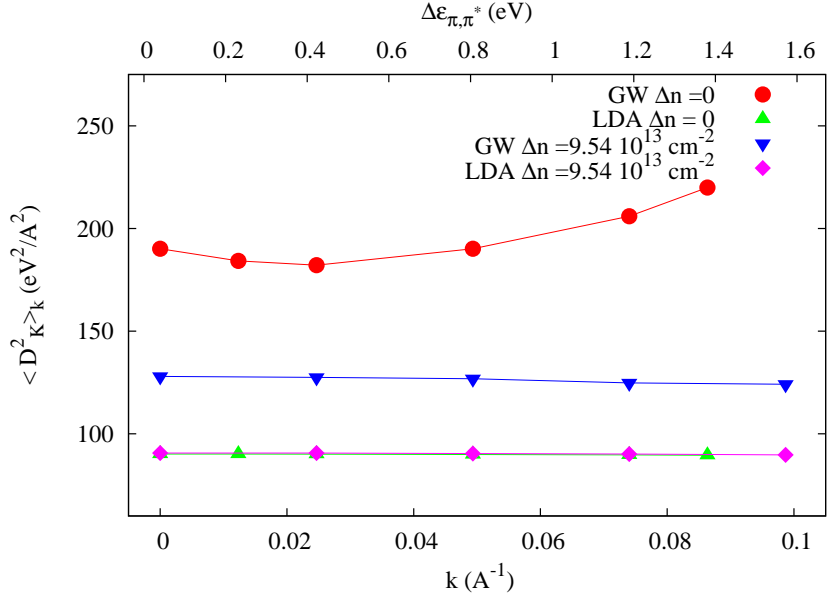}}
\caption{ Square of the deformation potential,
$\langle D^2_{\bf K}\rangle^{\pi\pi^*}_{\bf k}$,
with {\bf k} varying along the line ${\bf \Gamma}-{\bf M}$, perpendicular to ${\bf \Gamma}-{\bf K}$.
Four cases are reported: circles and up triangles are GW and LDA results for the undoped case; down-triangles and rhombus are GW and LDA for high doping $\Delta n=9.54 10^{13} cm^{-2}$.
The upper labels refere to the corresponding electronic gap
(the energy difference between $\pi^*$ and $\pi$ bands at {\bf K+k} according to GW calculations at zero doping).
\label{figepcline}}
\end{figure}

\begin{figure}[h]
\centerline{\includegraphics[width=0.5\textwidth,angle=-90]{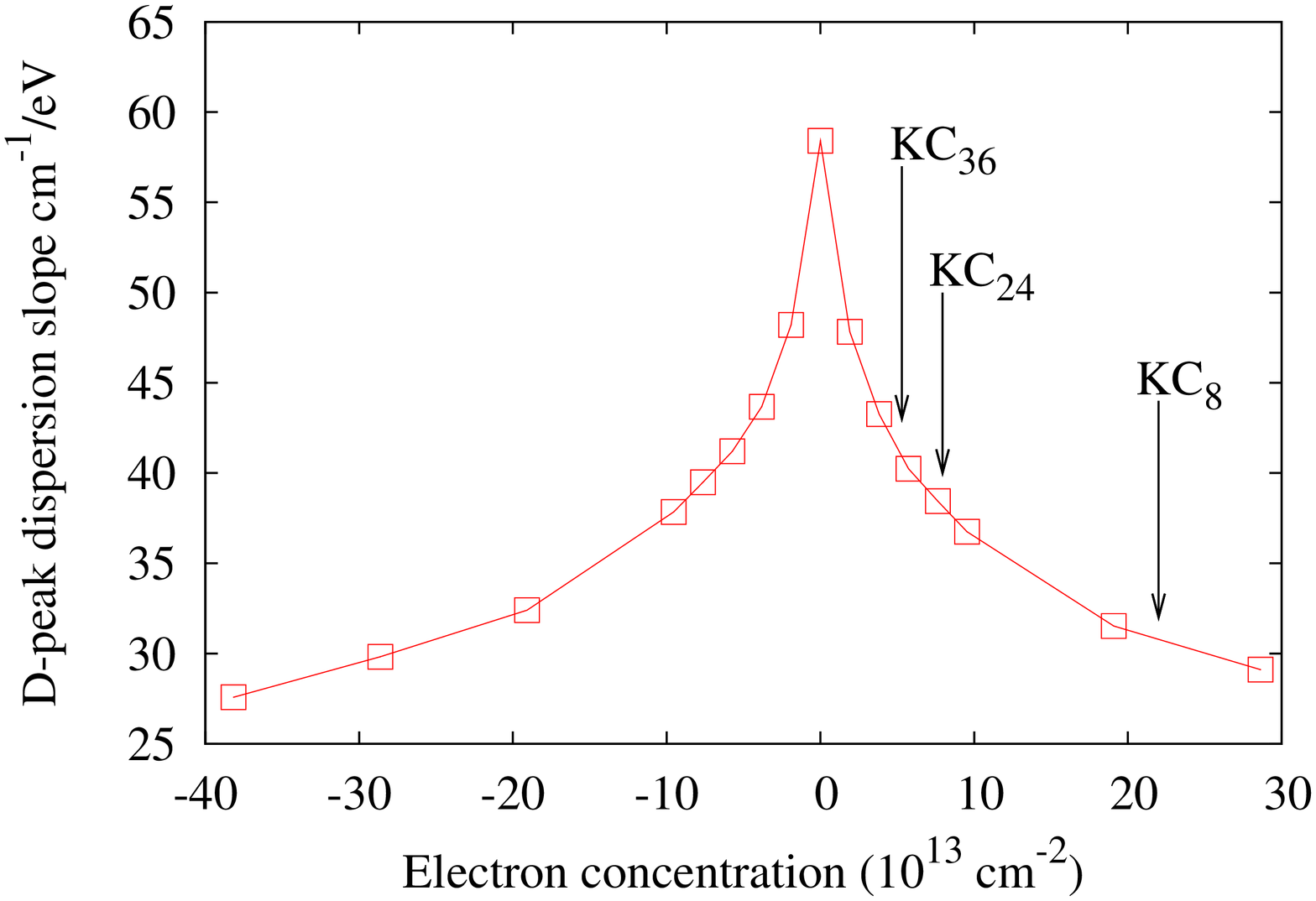}}
\caption{Change of slope of the Raman D-peak dispersion versus doping \label{fig3}.
Arrows indicate the equivalent doping level for the KC$_8$, KC$_{24}$, KC$_{36}$ intercalated graphite.}
\end{figure}

At this point it is instructive to check the performance of DFT with
hybrid functionals. We have performed calculations with the B3LYP
functional\cite{b3lypfunctional,b3lyp}.
Apart from an overestimation of $\langle D^2_{\bf K}\rangle^{\pi\pi^*}_{\bf k=0}$
by about 25\% (which could be corrected for by diminishing the percentage
of Hartree-Fock exchange in the functional), the calculation 
reflects rather the doping dependence of the GW-calculation with 
constant screening than the dependence of the full GW-calculation.
This can be understood, because the B3LYP functional contains screening
on a very simplified level not suitable to describe extended systems and metals.\\
Now we want show how the variation of the deformation potential affects the Raman D peak dispersion and the splitting of the 2D line in multilayer graphene. The dispersion (peak energy versus laser frequency) of the Raman D and 
2D lines in graphene is conveniently described by the double-resonant Raman model\cite{drr}. 
In order to calculate the phonon dispersion we need 
the deformation potential for a phonon wavevector {\bf K+q}
($\langle D^2_{\bf K+q}\rangle^{\pi\pi^*}_{\bf k}$).
In our earlier works we assumed this as a constant in {\bf q} and {\bf k}.
The calculation for a finite {\bf q} requires the use of very large
supercells, which are challenging for the GW approach.
On the contrary, we can test the dependence of the deformation potential on the electron
wavevector {\bf k}.
Indeed, we calculated $\langle D^2_{\bf K}\rangle^{\pi\pi^*}_{\bf k}$
for {\bf k} varying along the line ${\bf \Gamma}-{\bf M}$\cite{epcline}.
We found that $\langle D^2_{\bf K}\rangle^{\pi\pi^*}_{\bf k}$  is almost constant in the doped case and
it is slightly varying in the undoped one (within $10\%$), the result is reported in fig.~\ref{figepcline}.
This finding justifies our previous assumption\cite{lazzeri08} of a constant square deformation potential that was also verified by direct comparison with the experiments\cite{alexnew}. In the   double-resonant Raman model\cite{drr} the D-line dispersion is proportional to the phonon slope around {\bf K} and thus proportional to $\langle D^2_{\bf K}\rangle^{\pi\pi^*}_{\bf k}$.
Furthermore, it is inversely proportional to the slope of the
$\pi$/$\pi^*$ bands and thus inversely proportional to $\Delta \epsilon_m$.
In our previous work using the result at zero doping, we were able to 
reproduce completely \emph{ab-initio} the Raman D-peak \cite{lazzeri08} 
dispersion (peak position versus laser frequency). 
With the information on the doping dependence of 
$\langle D^2_{\bf K}\rangle^{\pi\pi^*}_{\bf k}$ and of $\Delta \epsilon_m$, 
we have calculated the Raman D-peak dispersion as a function of the doping with the approach described in ref.~\onlinecite{lazzeri08}. 
In Fig.~\ref{fig3} we report the resulting slope of the D-peak dispersion
obtained as a linear fit of the dispersion between 1.0 and 3.2 eV
laser energy.  The D-peak dispersion is almost symmetric with respect to electron/hole doping, and it has its maximum at zero doping.  Due to this strong variation with doping it can be used also to detect experimentally the charge state of a graphene sample.  
Finally we suggest a simple way to verify our result in multi-layer graphene using a single laser\cite{multilayer}. 
In multilayer graphene the 2D peak splits in different sub-peaks and this splitting is proportional to the D-peak dispersion with the laser frequency \cite{ferrari06,graf07}.
Therefore we expect that measurements of the 2D peak splitting as function of doping can highlight the strong variation of the D-peak dispersion, due to the squared deformation potential, as predicted in this work.

%Moreover, the ratio 
%$\langle D^2_{\bf \Gamma}\rangle_{\rm F}/\langle D^2_{\bf K}\rangle_{\rm F}$ 
%determines the doping dependence of the ratio between the intensities of the 
%G and D Raman peaks (see Ref~\onlinecite{baskonew}).

In conclusion, we have shown that in graphene 
$\langle D^2_{\bf K}\rangle^{\pi\pi^*}_{\bf k}$ can be easily tuned with doping. 
This means that the electron and phonon interaction cannot be 
described by a simplified  Hamiltonian with a fixed deformation potential 
as it is commonly done.
Our findings can be easily verified experimentally by measuring the doping dependence of 
the Raman D-peak dispersion, of the splitting of the 2D peak in multi-layer graphene 
and of the slope of the highest phonon branch close to {\bf K}.
The present findings open the possibility to use the slope of the D-peak dispersion
as a simple probe for the electrons/holes doping in graphene. 
The fact that the deformation potential is not a constant has implications also 
in the realization of graphene-based electronic devices where it is known 
that one of the limitation for ballistic transport is just the coupling 
between electrons and phonons\cite{zao,mauritube,highgraphene}. Tuning electron-phonon coupling 
by doping could boost electronic properties of graphene based devices.
Concerning the puzzling discussion on the size of the EPC in graphene, 
our result puts another piece in support that  EPC has to be larger 
than the LDA one and doping dependent.
Finally, this new way to tune electron-phonon coupling can play a fundamental role in all the experiments where physical phenomena are directly related to doping as for instance the phonon renormalization\cite{geim}, phonon linewitdh\cite{giustino2},  or the radiative decay of excitons in nanotubes\cite{perebeinos}. 

We acknowledge funding by the Spanish MEC (FIS2007-65702-C02-01), "Grupos Consolidados UPV/EHU del Gobierno Vasco" (IT-319-07), the European Community through  e-I3 ETSF project (Contract Number 211956), the french ANR project ACCATTONE. We acknowledge support by the Barcelona Supercomputing Center, "Red Espanola de Supercomputacion", SGIker ARINA (UPV/EHU) and Transnational Access Programme HPC-Europe++.


\begin{thebibliography}{99}
\bibitem{geim-rmp}
A. H. Castro Neto, F. Guinea,N. M. R. Peres,K. S. Novoselov and A. K. Geim, Rev. Mod. Phys. {\bf 81}, 109 (2009) 

\bibitem{tsang}
Tsang, {\it et al.}, Nature nanotechnology {\bf 2}, 725 (2007)

\bibitem{zao}
Z. Yao {\it et al.}, Phys. Rev. Lett. {\bf 84}, 2941 (2000); A. Javey {\it et al.}, Phys. Rev. Lett. {\bf 92}, 106804 (2004) 

\bibitem{mauritube}
M. Lazzeri, S. Piscanec, F. Mauri, A. C. Ferrari, and J. Robertson, Phys. Rev. Lett. {\bf 95}, 236802 (2005),
B. J. LeRoy, S. G. Lemay, J. Kong, and C. Dekker, Nature (London) {\bf 432}, 371 (2004)

\bibitem{highgraphene}
A. Barreiro, M. Lazzeri, J. Moser, F. Mauri, and A. Bachtold, Phys. Rev. Lett. {\bf 103}, 076601 (2009)

\bibitem{geim}
A. B. Kuzmenko, L. Benfatto, E. Cappelluti, I. Crassee, D. van der Marel, P. Blake, K. S. Novoselov, Phys. Rev. Lett. {\bf 103}, 116804 (2009)


\bibitem{alexkc8}
A. Gr\"uneis, C. Attaccalite, A. Rubio, D. V. Vyalikh, S. L. Molodtsov, J. Fink, R. Follath, W. Eberhardt, B. B\"uchner, and T. Pichler, Phys. Rev. B {\bf 79}, 205106 (2009)

\bibitem{alexnew}
A. Gr\"uneis, J. Serrano, A. Bosak, M. Lazzeri, S. L. Molodtsov, L. Wirtz, C. Attaccalite, M. Krisch, A. Rubio, F. Mauri, and T. Pichler, Phys. Rev. B {\bf 80}, 085423 (2009) 

\bibitem{andrei}
Guohong Li, Adina Luican and Eva Y. Andrei, Phys. Rev. Lett. {\bf 102}, 176804 (2009)

\bibitem{raman}
A. Das, {\it et al.}, Nature Nano {\bf 3}, 210 (2008), A. Das, Phys. Rev. B {\bf 79}, 155417 (2009)
%C. V. Raman, K. S. Krishnanm, Nature {\bf 121}, 501 (1928)

\bibitem{ferrari06}
A.C. Ferrari, J.C. Meyer, V. Scardaci, C. Casiraghi, M. Lazzeri, F. Mauri, 
S. Piscanec, D. Jiang, K. S. Novoselov, S. Roth, and A. K. Geim,
Phys. Rev. Lett. {\bf 97}, 187401 (2006)

\bibitem{graf07}
D. Graf, F. Molitor, K. Ensslin, C. Stampfer, A. Jungen, C. Hierold, 
and L. Wirtz, Nano Lett. {\bf 7}, 238 (2007) 

\bibitem{pis07}
S. Pisana, M. Lazzeri, C. Casiraghi, K. S. Novoselov, A. K. Geim, 
A. C. Ferrari, and F. Mauri, Nature Mat. {\bf 6}, 198 (2007).  

\bibitem{efield}
J. Yan, Y. Zhang, P. Kim, and A. Pinczuk, Phys. Rev. Lett. {\bf 98}, 166802 (2007) 

\bibitem{ferrari07b}
C. Casiraghi, S. Pisana, K.S. Novoselov, A.K. Geim, and A.C. Ferrari,
Appl. Phys. Lett. {\bf 91}, 233108 (2007). 
	

\bibitem{stampfer07}
C. Stampfer, F. Molitor, D. Graf, K. Ensslin, A. Jungen, C. Hierold, and
L. Wirtz, Appl. Phys. Lett. {\bf 91}, 241907 (2007) 

\bibitem{baskonew}
D. M. Basko, S. Piscanec, A. C. Ferrari, Phys. Rev. B {\bf 80}, 165413 (2009) 

\bibitem{gpeakdoping}
M. Lazzeri and F. Mauri, Phys. Rev. Lett. {\bf 97}, 266407 (2006)

\bibitem{baskoedge}
C. Casiraghi, A. Hartschuh, H. Qian, S. Piscanec, C. Georgi, K. S. Novoselov, D. M. Basko, A. C. Ferrari, 	Nano Lett., {\bf 9}, 1433 (2009)


\bibitem{calandra}
Matteo Calandra and Francesco Mauri, Phys. Rev. B {\bf 76}, 205411 (2007) 

\bibitem{drr}
C. Thomsen and S. Reich,  Phys. Rev. Lett. {\bf 85}, 5214 (2000).

\bibitem{epcline}
We consider {\bf k} along the line ${\bf \Gamma}-{\bf M}$,
perpendicular to ${\bf \Gamma}-{\bf K}$, of the graphene unit-cell.
$\langle D^2_{\bf K}\rangle^{\pi\pi^*}_{\bf k}$ can be obtained from the
band energies of a $\sqrt{3}\times\sqrt{3}$ supercell (see Fig 3b of Ref.\cite{lazzeri08}).
The two Dirac cones at {\bf K} and 2{\bf K} of the unit-cell
are refolded at ${\bf \Gamma}$ in the supercell.
We dispalce each atom by $d$, following the {\bf K}-A'$_1$ phonon pattern. 
We define $\Delta \epsilon_d({\bf k})={\overline \epsilon_{\pi^*}}({\bf k})-{\overline \epsilon_{\pi}}({\bf k})$,
where ${\overline \epsilon_{\pi^*}}$ (${\overline \epsilon_{\pi}}$) is the
average between the energy of the two $\pi^*$($\pi$) bands corresponding to 
{\bf K}+{\bf k} and 2{\bf K}+{\bf k} of the unit-cell $\langle D^2_{\bf K}\rangle^{\pi\pi^*}_{\bf k}=((\Delta \epsilon_d({\bf k}))^2-(\Delta \epsilon_0({\bf k}))^2
)/(8d^2)$.

\bibitem{piscanec04}
S. Piscanec, M. Lazzeri, F. Mauri, A.C. Ferrari, and J. Robertson,
Phys. Rev. Lett. {\bf 93}, 185503 (2004).                     

\bibitem{ando}
K. Ishikawa and T. Ando, J. Phys. Soc. of Japan, {\bf 75}, 084713 (2006)

\bibitem{saha}
S.K. Saha, U.V. Waghmare, H.R. Krishnamurthy, and A.K. Sood,
Phys. Rev. B {\bf 76}, 201404 (2007).

\bibitem{dassarma}
W. K. Tse and S. Das Sarma, Phys. Rev. Lett. {\bf 99}, 236802 (2007)

\bibitem{lazzeri08}
M. Lazzeri, C. Attaccalite , L. Wirtz , and F. Mauri
Phys. Rev. B {\bf 78}, 081406(R) (2008) 

\bibitem{basko}
D. M. Basko and I. L. Aleiner, Phys. Rev. B {\bf 77}, 041409(R) (2008).

\bibitem{kc8} 
A. Gr\"uneis, C. Attaccalite, A. Rubio, D. V. Vyalikh, S. L. Molodtsov, J. Fink, R. Follath, W. Eberhardt, B. B\"uchner, and T. Pichler, Phys. Rev. B {\bf 80}, 075431 (2009) 

%\bibitem{lazzeri06b}
%M. Lazzeri and F. Mauri, Phys. Rev. B {\bf 73}, 165419 (2006)


\bibitem{calculation}
 In all DFT calculation the distance between the graphene planes was 20 a.u., the Brillouin Zone integration  was performed using an uniform k-point grid $36 x 36 x 1$, with the functional of Ref.~\onlinecite{ceperley}, plane waves (60 Ry cut-off) and pseudo-potentials \cite{troullier}, using  the PWSCF code \cite{pwscf}. An electronic smearing of $0.02$ Ry with the Fermi-Dirac distribution was employed.

\bibitem{laz06}
M. Lazzeri, F. Mauri Phys. Rev. Lett. {\bf 97}, 266407 (2006)

\bibitem{louie}
C. H. Park, F. Giustino, M. L. Cohen and S. G.. Louie, Phys. Rev. Lett. {\bf 99}, 086804 (2007)

\bibitem{tbarpes}
A. Gr\"uneis et al., Phys. Rev. B, {\bf 78}, 205425 (2008) 


\bibitem{GW}
Non-self consistent GW calculations have been performed 
starting from DFT-LDA wave-functions, using a plasmon pole approximation, following the
scheme of Hybertsen and Louie\cite{hybert}, with  the code YAMBO\cite{yambo}.
We use a 36x36x1 k-point grid for the primitive cell and an equivalent one for the supercell. Convergence in the number of bands and size of the dielectric constant has been carefully  checked.

\bibitem{hybert}
M.S. Hybertsen and S.G. Louie, Phys. Rev. B {\bf 34}, 5390 (1986).


\bibitem{zhou}
S.Y. Zhou et al., 	Nature Phys. {\bf 2}, 595-599 (2006)

\bibitem{graphite}
A. Gr\"uneis et al., Phys. Rev. Lett. {\bf 100}, 037601 (2008)

\bibitem{nonadiabatic}
Notice that non-adiabatic corrections, not considered here, are though to be important for the G phonon but not for the {\bf K} phonon, see Ref.~\onlinecite{laz06}

\bibitem{b3lypfunctional}
A.D. Becke, J. Chem. Phys. {\bf 98}, 5648 (1993).

\bibitem{b3lyp}
For the B3LYP calculations we used the code {\tt CRYSTAL}
(V.R. Saunders et al., R. Dovesi, C. Roetti, R. Orlando, C.M. Zicovich-Wilson,
N.M. Harrison, K. Doll, B. Civalleri, I.J. Bush, Ph. D’Arco, M. Llunell
{\tt CRYSTAL03} User’s Manual, University of Torino, Torino, 2003),
using the TZ basis by Dunning (without the diffuse P-function).
K-point sampling and thermal smearing are the same as in the GW-calculations.



\bibitem{polini}
M. Polini, R. Asgari, Y. Barlas, T. Pereg-Barnea, A.H. MacDonald, Solid State Communications {\bf 143}, 58 (2007)  

\bibitem{attaccalite}
C. Attaccalite, A. Rubio, Physica Status Solidi B {\bf 246}, 2523 (2009)

\bibitem{multilayer}
We suppose that also in multi-layer graphene, althougt the band structure it is different from graphene, a result similar to the one we found for graphen holds, as we found for graphite\cite{lazzeri08}.

\bibitem{magnetic}
C. Faugeras, M. Amado, P. Kossacki, M. Orlita, M. Sprinkle, C. Berger, W.A. de Heer, M. Potemski, Phys. Rev. Lett. {\bf 103}, 186803, (2009)


%\bibitem{saitta}
%A. M. Saitta, M. Lazzeri, M. Calandra, F. Mauri,  Phys Rev Lett. {\bf 100}, 226401 (2008) 

%\bibitem{yan}
%J. Yan, Y. Zhang, P. Kim and A. Pinczuk, Phys Rev Lett. {\bf 98}, 166802 (2007) 


%\bibitem{tsung}
%Tsung-Ta Tang {\it et al.}, http://arxiv.org/abs/0907.0419







%\bibitem{basko08}
%D. M. Basko, Phys. Rev. B 78, 125418 (2008)




\bibitem{giustino2}
C. H. Park, F. Giustino, M. L. Cohen, S. G. Louie, Nano Lett. {\bf 8}, 4229 (2008)

\bibitem{perebeinos}
V. Perebeinos and P. Avouris, Phys. Rev. Lett. {\bf 101}, 057401 (2008)

\bibitem{ceperley} 
D.M. Ceperley and B.J. Alder, Phys. Rev. Lett. {\bf 45}, 566 (1980)

\bibitem{troullier} 
N. Troullier and J. L. Martins, Phys. Rev. B {\bf 43}, 1993 (1991)
           
\bibitem{pwscf}
P. Giannozzi et al. J. Phys.: Condens. Matter {\bf 21}, 395502 (2009), http://www.quantum-espresso.org



%\bibitem{deformation}
%J. Bardeen and W. Shockley, Phys. Rev.  {\bf 80}, 72 (1950).


%\bibitem{weshown}
%We showed in a previous work, Ref.~\onlinecite{lazzeri08}, that this scheme introduces a large renomrlaisation also in the phonon structure and in particular for the $K$ phonon frequency, changing the dependence of the Raman D line with laser frequency in agreement with experiments.

\bibitem{yambo}
A. Marini, C. Hogan, M. Gruning, and D. Varsano, Comp. Phys. Comm. {\bf 180}, 1392 (2009), http://www.yambo-code.org.


%\bibitem{DFPT}
%Electron-Phonon calculations were done using the method described in 
%Ref.~\onlinecite{baroni} using the LDA approximation.
%Technical details are the same as in Ref.~\onlinecite{piscanec04}.
%
%\bibitem{baroni}
%S. Baroni {\it et al.}, Rev.\ Mod.\ Phys.\ {\bf 73}, 515 (2001),



%\bibitem{pisana}
%S. Pisana et al. Nature Mat. {\bf 6},198 (2007)

%\bibitem{das}
%A. Das,S. Pisana, B. Chakraborty, S. Piscanec, S. K. Saha,U. V. Waghmare, K. S. Novoselov, H. R. Krishnamurthy, A. K. Geim,  A. C.  Ferrari, A. K. Sood,  Nat. Nanotechnol. {\bf 3}, 210 – 215 (2008)

%\bibitem{casiraghi07}
%C. Casiraghi, S. Pisana, K. S. Novoselov, A. K. Geim, A. C. Ferrari, Appl. Phys. Lett., {\bf 91}, 233108 (2007)

%\bibitem{casiraghi08}
%C. Casiraghi, A. Hartschuh, H. Qian, S. Piscanec, C. Georgi, K. S. Novoselov, D. M. Basko, A. C. Ferrari, http://arxiv.org/abs/0810.5358





\end{thebibliography}
\end{document}